\documentclass[sigconf, authorversion]{acmart}
\AtBeginDocument{%
  \providecommand\BibTeX{{%
    \normalfont B\kern-0.5em{\scshape i\kern-0.25em b}\kern-0.8em\TeX}}}

\copyrightyear{2021} 
\acmYear{2021} 
\setcopyright{acmlicensed}
\acmDOI{10.1145/3411764.3445663}

\acmConference[CHI '21]{CHI Conference on Human Factors in Computing Systems}{May 8--13, 2021}{Yokohama, Japan}
\acmBooktitle{CHI Conference on Human Factors in Computing Systems (CHI '21), May 8--13, 2021, Yokohama, Japan}
\acmPrice{15.00}

\definecolor{darkgreen}{rgb}{0, 0.5, 0}

\newcommand*{\eg}{e.g., {}}
\newcommand*{\ie}{i.e., {}}

\makeatletter
\newcommand*{\etal}{%
    \@ifnextchar{'}%
        {et~al}%
        {et~al. {}}%
}

\newcommand{\specialcell}[2][c]{%
  \begin{tabular}[#1]{@{}c@{}}#2\end{tabular}}

\begin{document}

% Original submission title
%\title{It's Not the Movement: The Central Information Needed in Stroke Telerehabilitation}
%\title{It's Not Just About the Movement: Experiential Information Needed in Stroke Rehabilitation}
\title{It's Not Just the Movement: Experiential Information Needed for Stroke Telerehabilitation}
\renewcommand{\shorttitle}{Experiential Information Needed for Stroke Telerehabilitation}


\author{Adegboyega Akinsiku}
\affiliation{%
 \institution{University of Maryland, Baltimore County}
 \city{Baltimore}
 \state{MD}
 \country{USA}}
   \email{aakins1@umbc.edu}
    
\author{Ignacio Avellino}
\affiliation{%
 \institution{University of Maryland, Baltimore County}
 \city{Baltimore}
 \state{MD}
 \country{USA}}
 \email{ignacio.avellino@umbc.edu}

\author{Yasmin Graham}
\affiliation{%
 \institution{University of Maryland, Baltimore County}
 \city{Baltimore}
 \state{MD}
 \country{USA}}
  \email{ygraham@berkeley.edu}
  
  \author{Helena M. Mentis}
\affiliation{%
 \institution{University of Maryland, Baltimore County}
 \city{Baltimore}
 \state{MD}
 \country{USA}}
  \email{mentis@umbc.edu}

\renewcommand{\shortauthors}{Akinsiku, Avellino, et al.}

\begin{abstract}
Telerehabilitation systems for stroke survivors have been predominantly designed to measure and quantify movement in order to guide and encourage rehabilitation regular exercises at home. We set out to study what aspect of the movement data was essential, to better inform sensor design. We investigated face-to-face stroke rehabilitation sessions through a series of interviews and observations involving 16 stroke rehabilitation specialists including physiatrists, physical therapists, and occupational therapists. We found that specialists are not solely interested in movement data, and that experiential information about stroke survivors' lived experience plays an essential role in specialists interpreting movement data and creating a rehabilitation plan. We argue for a reconceptualization in stroke telerehabilitation that is more inclusive of non-movement data, and present design implications to better account for experiential information in telerehabilitation systems.
\end{abstract}

\begin{CCSXML}
<ccs2012>
   <concept>
       <concept_id>10003120.10003121.10011748</concept_id>
       <concept_desc>Human-centered computing~Empirical studies in HCI</concept_desc>
       <concept_significance>500</concept_significance>
    </concept>
</ccs2012>
\end{CCSXML}
\ccsdesc[500]{Human-centered computing~Empirical studies in HCI}


\keywords{Telerehabilitation; Stroke; Physical medicine and rehabilitation; Therapy}

\maketitle

\newpage
% ==================== ====================
\section{Introduction}
% ==================== ====================

Telerehabilitation involves leveraging technologies (\eg the Internet) to facilitate \textit{the communication of information} between a patient and their clinician at a distance in order to provide rehabilitation services~\cite{Brennan2010}. However, the type of information that needs to be communicated is not well defined. Prior to the viability of telerehabilitation services, researchers and developers designed non-internet \emph{home-based} therapy systems with the intention of increasing patients' access and longevity to rehabilitation services. These home-based therapy systems largely focused on motivating exercise and creating engagement (\eg through gamification) at home. As telerehabilitation systems' viability and development increased, they seem to have organically followed the focus of home-based therapy technologies, where patients autonomously perform exercises, and movement data was primarily captured and conveyed to rehabilitation specialists. Thus, research and development both in telerehabilitation and home-based systems have largely aimed at sensing movement data.

Our research was initially guided by this focus on collecting and sharing movement data. We set out to determine the types and exactness of movement data needed in stroke rehabilitation, with the goal of informing the design of telerehabilitation systems for low-resource communities. As we began to observe face-to-face rehabilitation sessions and interview physiatrists, physical therapists, and occupational therapists, we instead found that the information that they dedicated effort to extract, understand, and integrate into their care plans, is incongruent with the current design paradigm of telerehabilitation systems.

In this study, we investigate the information exchanged between stroke survivors, clinicians, and caregivers in co-located in-clinic stroke rehabilitation sessions, with the goal of informing the design of future stroke telerehabilitation systems, such that patients and specialists can attain the benefits of co-located interaction. What we learned is that the information needed by rehabilitation specialists is not really a detailed understanding of the movement data, but rather a deep understanding of the \emph{experiential information}, such as the stroke survivor's emotions and motivations. We show that experiential information is information shared in stroke rehabilitation. Therefore, we posit for a paradigm reconceptualization for telerehabilitation system design, in which telerehabilitation has a focus on communicating the patient's situated context. 
\newpage
Our contributions to HCI research on stroke telerehabilitation are: (1) The definition and composition of the \emph{experiential information} needed in stroke rehabilitation, (2) An explanation of our proposed paradigm reconceptualization for future stroke telerehabilitation research, and, (3) Implications for the design and development of stroke telerehabilitation systems.

% ==================== ====================
\section{Background: Stroke Rehabilitation Process \& Specialists}
%\label{sec:background}
% ==================== ====================
Stroke is one of the leading causes of long-term disability in the United States~\cite{benjamin2017heart}. The American Heart Association and American Stroke Association projected in 2013 that the costs associated with stroke will increase 129\% by 2030, and concluded that more rehabilitation and acute care services are needed to address stroke because of the national healthcare costs increasing yearly~\cite{Ovbiagele2013}. Stroke survivors in particular, will spend directly and indirectly an average of $\$103,576$ over their lifetime on treatment~\cite{Heidenreich2011}. Rehabilitation after a stroke has a high cost as it involves a wide variety of experienced clinicians and specialized equipment. Unfortunately, access to specialized rehabilitation locations puts high-level care outside of the reach of many US citizens, including those in rural and low-resource communities~\cite{benjamin2017heart}. The ongoing COVID-19 pandemic has exacerbated these challenges, as today remote healthcare is the only treatment option even for people where distance and cost was not a barrier.

To give background, we describe the stroke rehabilitation process through the perspective of the multiple rehabilitation specialists that create rehabilitation care plans for outpatient stroke survivors. These specialists coordinate through an extensive care network that includes: (1) the stroke survivor, (2) caregivers (\ie the stroke survivor's immediate care network), (3) medical specialists (\eg physiatrists, cardiologists, and neurologists), and (4) allied health specialists (physical therapists, occupational therapists, and speech-language therapists). The network engages in an extensive amount of co-interpretation~\cite{Mentis2015}, a collaborative and interpretive process to assess movement and treatment efficacy, and care coordination~\cite{McDonald2007}, organizing the different aspects of care.

\textbf{\textsc{Physiatrists (PHY).}}
Physical Medicine and Rehabilitation (PM\&R), or physiatry, is the branch of medicine that treats individuals with physical impairments, functional limitations, pain, and disabilities that affect the brain, spinal cord, nerves, bones, joints, ligaments, muscles, and tendons~\cite{nih-pmr-clinical-center_2017, AmericanAcademyofPhysicalMedicineandRehabilitation2020}. Physiatrists are the primary medical doctors that guide stroke rehabilitation treatment, preceding physical and occupational therapy. They are key for acquiring a holistic perspective on functional and motor stroke rehabilitation, even if, to our knowledge, no other qualitative HCI paper has included physiatrists in their studies.

\textbf{\textsc{Physical Therapist (PT) and Occupational Therapist (OT).}}
A PT provides care to restore and maintain a patient's sensory and motor abilities (\eg improve gross motor movement), whereas an OT provides care to reduce the effects of a patient's disabilities through adaptation (\eg retrain self-care skills)~\cite{nih-ot-clinical-center_2017,nih-pt-clinical-center_2019}. Both specialists have aligned objectives in providing goal-oriented care to their patients, to ultimately restore functional ability and mobility, and improve the quality of life through a \emph{patient-centered approach}.

\textbf{\textsc{Patient-Centered Care.}} Patient-centered care takes into account the individual needs, values, and expressed interest of patients; and it has been identified as a gap in the US health system, with the Institute of Medicine urging the United States Congress to establish funds for this purpose~\cite{InstituteofMedicine2001}. Six primary dimensions make up patient centered care~\cite{gerteis1993}: (1) respect for patients' values, (2) coordination and integrative care, (3) information, communication, and education, (4) physical comfort, (5) emotional support to combat fear and anxiety, and (6) involvement of patients' family and friends. Patient-centered care is enacted in face-to-face stroke rehabilitation sessions by specialists assessing the patient's rehabilitation progress, having their patients conduct interventions (exercises and activities), and then create or modify an existing rehabilitation care plan in concert with the patient. The care plan is typically a set of prescribed interventions that the specialists evaluate and update periodically to monitor progress, making it dynamic and evolving. Specialists typically document their assessment using the SOAP method (Subjective, Objective, Assessment, and Plan), a method widely used by healthcare professionals to fill out patient notes during an appointment, and used to promote continuity in health records~\cite{Weed1964}.

% ==================== ====================
\section{Related Work on Home-based and Telerehabilitation System Design}
% ==================== ====================
Technology for stroke rehabilitation can be classified as \emph{home-based therapy systems}, which are not connected online to remote specialists, and \emph{telerehabilitation systems}, which communicates rehabilitation information and data through the Internet to a remote specialist~\cite{Brennan2010}. Telerehabilitation systems vary in implementation, but they focus on, asynchronously or synchronously, connecting patients to remote specialists and transmit: (1) communication data (\eg audio, video or text message) and/or (2) sensor-based data (\eg movement).

After reviewing systematic and scoping reviews of telerehabilitation systems from the last 10 years~\cite{Johansson2011,Santayayon2012,Laver2013,Rogante2015,Veras2017,Sarfo2018,Tchero2018,Appleby2019}, we noticed a trend that current stroke telerehabilitation system design has centralized the asynchronous and synchronous communication of movement sensor data. This trend has its own complexities outside the scope of this work, and deserves to be studied on its own in the future.
The following review of home-based and telerehabilitation design below exposes a clear trend of focusing on movement data, typically originating from sensors, rather than considering the importance of other information needs and goals.
Most of this research is driven by generating solutions that increase exercise motivation.
% ========== ==========
\subsection{Sensors and Motivation as the Design Focus}
% ========== ==========
% this is also about virtual training
Data needs in stroke rehabilitation systems had been codified in 2009 in Egglestone \etal's~\cite{Egglestone2009} design framework for home-based stroke therapy systems. Through workshops with stroke survivors and clinicians, the authors identified (1) background information, such as the stroke's disruptive effects on patients or the contribution of professionals to recovery, and (2) exercise execution data for evaluation, that can be gathered through sensors or self-report. What has transpired since then is a litany of sensor-based systems focused on gathering data to support the second information need. What we have not seen is a consideration of the first form of information need. Specifically, identifying the types of ``background'' information and how to collect the information in way to present the two types of information to rehabilitation specialists.

Sensor development has been an important step for telerehabilitation and home-based systems to work---we are not denying that. However, consider the following five recently published systems:
a low-cost, wireless home-based rehabilitation sensor that reliably captures upper-limb arm posture and movement~\cite{Lim2010};
\emph{Us'em}~\cite{Beursgens:2011:UMS:1979742.1979761}, a wristband-like activity monitor of arm--hand performance designed strictly for patients to motivate the use of an impaired arm during everyday activities;
\emph{mRes}~\cite{Weiss:2014:LCT:2686893.2686989}, a low-cost device that measures rotational movement, aimed at training dorsal wrist extension and finger manipulation (both in supination and pronation), with an API for information exchange with telerehabilitation systems;
The combination of Microsoft's Kinect\footnote{\url{https://developer.microsoft.com/en-us/windows/kinect/}} sensor data with machine learning to automatically assess stroke rehabilitation exercises~\cite{lee2019}; and,
\emph{ArmSleeve}~\cite{Ploderer2016}, a sensor-embedded sleeve that captures objective upper limb data in patients' daily life, outside rehabilitation exercises, creating a visualization for OTs in a dashboard.
In all of these, the initial focus is on the valid collection of data towards motivating correct movement. Interestingly, in the evaluation of this last system, the authors found a major limitation to interpreting this movement data: the lack of contextual information.

Motivating a patient to perform, or more so ``correctly'' perform, a movement or exercise has been a prevailing goal in many of the systems designed. For instance, Alankus \etal~\cite{Alankus2010} studied therapeutic games using Nintendo Wii remote controllers\footnote{\url{https://www.nintendo.co.uk/Wii/Accessories/Accessories-Wii-Nintendo-UK-626430.html}} and a webcam to sense movement, emphasizing the role of home-based stroke rehabilitation games in keeping monotony low while providing performance feedback to specialists. In their followup work, Alankus \etal~\cite{Alankus2012c} specifically used the Wii remote to reliably sense compensatory movement (i.e. ``incorrect'' movements to achieve the same outcome because ``correct'' movement is not possible or tiring) during gameplay, and then recommend feedback mechanisms to reduce such compensatory movements. Likewise, mobile phone sensors have also been used for keeping rehabilitation engaging, measuring movement in upper-limb rehabilitation while providing instant feedback on the screen through a game~\cite{10.1145/2700648.2811337}.

We do not disagree that giving immediate feedback to patients about the correctness of exercise execution can help improve their performance as well as motivate them to try harder. For instance, vibrotactile feedback~\cite{Held2017} in an Arm Usage Coach (AUC), or using a Virtual Reality headset to be immersed in a 3D game that adapts its intensity to increase engagement ~\cite{10.1145/3316782.3321545}. What we are questioning with this paper is how useful is movement information in a telerehabilitation context and how might the focus on this design trend miss the information needs of rehabilitation specialists.

% ========== ==========
\subsection{Data Needs in Telerehabilitation}
% ========== ==========
In the development of telerehabilitation systems, there is added complexity on determining what data/information to transmit to a care specialist, and how to present it in a meaningful way. For instance, after usability testing of \emph{TeleREHA}, Perry \etal~\cite{Perry2011} found there are data needs in planning (configuration, game parameterizing and scheduling), executing (measuring exercise data), and assessing (viewing data). In response, Postolache \etal~\cite{Postolache2011} proposed intelligent telerehabilitation assistants called \emph{Rehabilitative TeleHealthCare}, where sensor signals are processed and combined into visualizations for caregivers, including physical movement (\eg posture and daily walking movement), physiological data (\eg heart rate, oxygen saturation and respiratory rate) and localization.

However, there have been indications that movement data is not enough in the telerehabilitation context. Dekker-van~Weering \etal ~\cite{Dekker-vanWeering:2015:DTS:2838944.2838971} evaluated clinician needs using a telerehabilitation system and pointed to the system's need to integrate patient context. For instance, integrating patient's mood to allow them to skip a day of exercise, or engagement with a therapist when choosing exercises. Lastly, in telerehabilitation co-design sessions with traumatic brain injury clinicians, How \etal~\cite{How2017c} concluded that a successful system needs to adapt to a patient's physical, cognitive and emotional state, the evolution of their rehabilitation history, and the surrounding life context such as social life. Note that the authors argue that \emph{designers} should take into account these aspects when building systems, but they do not study in depth what this contextual information is or how its role in the system. 

In summary, stroke home-based therapy and telerehabilitation systems have followed a trend of collecting movement data. This trend leads to systems that provide little to no context of a patient's situated condition and excludes the collection of information like subjective data. In contrast, our research aims to integrate all components of face-to-face stroke rehabilitation into a telerehabilitation system design. We conducted a field study focused on face-to-face sessions to reveal information and practices that might not be integrated into existing telerehabilitation tools.

% ==================== ====================
\section{Method}
% ==================== ====================

% ========== ==========
\subsection{Field Study Design}
% ========== ==========
As we were interested in the the complex and rich information exchange that occurs in co-located rehabilitation, we conducted an ethnographically-informed field study involving rehabilitation sessions observations, interviews with specialists, and a review of stroke rehabilitation-related documents/artifacts. We held four consultation meetings with a neurologist and a PT, specialized in stroke rehabilitation and recovery, before starting the study to inform its design.

The semi-structured interviews involved specialists at the three different medical centers, and observations involved specialists who worked in two of them. Our initial observations were conducted in one hospital located in a major city in the mid-Atlantic region of the United States of America, later adding a hospital in a different system. All hospitals had Physical Medicine \& Rehabilitation departments. All of the medical centers served patients who were from (1) low socio-economic areas, (2) technologically low-resourced locations, (3) and/or surrounding rural areas. Documents and artifacts were collected at both hospitals as our study progressed.

% ========== ==========
\subsection{Participants}
% ========== ==========
Participants were recruited through snowball sampling within the participating medical centers. We interviewed four physiatrists (PHYs), five physical therapists (PTs) and seven occupational therapists (OTs), who work with stroke survivors in an outpatient setting, and observed a subset of those based on availability. The participant demographics are detailed in \autoref{tab:demoandobs}. Participants had various specialties, evident by their degrees (including M.D./Ph.D., Masters in Engineering, DPT, and MBA), as well as experience in different medical settings, most notably in low resource and rural communities outside of the US (denoted with ** in \autoref{tab:demoandobs}).

\begin{table}[h]
\smaller
\begin{tabular}{cccc}
\toprule
\multicolumn{1}{c}{Participant} & \specialcell{Experience\\ (years)} & \specialcell{Practice\\Setting}            & Data  \\
\midrule
PHY1                            & 15                 & P, H, R & I7, O9*,O10*    \\
PHY2                            & 17                 & P, H, R & I8  \\
PHY3**                           & 16                 & P, H           & I9   \\
PHY4                            & 6                  & H                    & I13   \\
PT1**                            & 21                 & P, H       & I4   \\
PT2                             & 29                 & P, H           & I10,O12*    \\
PT3                             & 3                  & H                    & I14,O18    \\
PT4                             & 5                  & H                    & I15,O13*,O14*,O15*,O16,O17  \\
PT5                             & 4                  & H                    & I16   \\
PT6                             & 7                  & H                    & O3\\
OT1                             & 5                  & H                    & I1,O4*,O5,O6,O7,O8\\
OT2                             & 18                 & P, H, R & I2   \\
OT3**                             & 2                  & H                    & I3   \\
OT4**                            & 6                  & P, H, R & I5   \\
OT5                             & 1y \& 8m       & P, H           & I6  \\
OT6**                             & 25                 & P, H           & I11,O11    \\
OT7                             & 2m           & P                      & I12   \\
OT8                             & 2                  & H                    & O1, O2\\
\bottomrule
\end{tabular}
\caption{Participant demographics and collected data summary. ** denotes experience outside of the US. * denotes caregiver was in attendance. ``I'' = Interview. ``O'' =  Observation. ``P'' =  Private. ``H'' =  Hospital. ``R'' =  Research.}
\label{tab:demoandobs}
\end{table}

% ========== ==========
\subsection{Data Collection}
% ========== ==========
% ===== =====
\subsubsection{Observations}
% ===== =====
We observed and video/audio recorded 18 stroke survivors rehabilitation sessions, to understand in-clinic rehabilitation, focusing on the information exchange between stroke survivors, caregivers, and rehabilitation specialists as they discuss rehabilitation.
Each session was video recorded, from the beginning to the end.
These sessions included specialists assessing the dexterity, spasticity, and cognitive function of stroke survivors, and subsequently, treatments, exercises, and therapies were prescribed.
The exercises and activities varied in the abilities they were attempting to address: motor, strength, cognitive, robotic, and aquatic.

Due to a COVID-19 state of emergency, we were unable to observe further physical therapy sessions, and we were not allowed to attend virtual sessions due to H policy.

% ===== =====
\subsubsection{Semi-Structured Interviews}
% ===== =====
We conducted semi-structured interviews with 16 of the 18 participants. We were unable to complete interviews with PT6 and OT8 due to workflow constraints before and after the observed rehabilitation sessions. For the participants we observed, interviews were conducted before and after the rehabilitation session at the convenience of the specialists. Before observations, interviews primarily focused on \textbf{(A)} gaining insight into the current work practices and data needs of the rehabilitation specialist and \textbf{(B)} eliciting their perceived needs for a telerehabilitation system. After observations, interviews focused on validating our interpretation of their practices and the information shared within the session. For those we were unable to observe, only one interview was conducted.

At the start of this study, the semi-structured interview protocol questions included:

\begin{enumerate}
    \item\textbf{A:} What are you trying to accomplish through the rehabilitation evaluations?
    \item\textbf{A:} What are you looking for when patients complete a task? (repetition, completion or form?) 
    \item\textbf{B:} What will you like to know about your patients when they complete activities at home? 
\end{enumerate}

Following the first set of observations and interviews with OT1 and OT2 and reflecting on our consultations, we realized the specialists' interests lay beyond information on repetition completion and form, as they provided a much broader set of information types. We thus added a new prompt, asking participants to rank four data types we had collected at that point according to importance. 

\begin{enumerate}
\setcounter{enumi}{3}
\item \textbf{B:}
 In the order of most importance to least, what data (activity repetition, frustration, motivation, and stress) are you interested in knowing about your patient's health at home while completing rehabilitation exercises?
\end{enumerate}

\begin{itemize}
\item \textit{Activity Repetition}: Completion of the prescribed repetitions, computed from sensor data.
\item \textit{Frustration}: Level of frustration when completing a prescribed exercise.
\item \textit{Motivation}: Level of motivation when completing a prescribed exercise. 
\item \textit{Stress}: Levels of stress due to stroke or home environment.
\end{itemize}

% ===== =====
\subsubsection{Documentation and Artifact Review}
% ===== =====
We collected documentation provided to stroke survivors to complement our understanding, including: a patient take-home exercise packet, a brochure for patients explaining stroke, a cognitive impairment assessment form called Montreal Cognitive Assessment (MOCA)\footnote{\url{https://www.mocatest.org/}}~\cite{Nasreddine2005}, and a medical note template that uses the \emph{Subjective, Objective, Assessment and Plan} (SOAP) method (example from~\cite{Susan2002} in~\autoref{app:SOAP}).
We also examined the \emph{Nine Hole Peg Test}, a quantitative assessment that measures finger dexterity\footnote{\url{https://www.sralab.org/rehabilitation-measures/nine-hole-peg-test}}, and a \emph{Box and Block test}, that asses unilateral gross movement\footnote{\url{https://www.sralab.org/rehabilitation-measures/box-and-block-test}}.

\newpage

% ========== ==========
\subsection{Ethical Considerations}
% ========== ==========
We obtained IRB approval from the University of Maryland, Baltimore County Institutional Review Board, and received approval to be on site from administrators at the partnering clinical institutions. We obtained consent from the clinician participants before the initial interview. Before observations began, we gained verbal consent from the patient and caregiver (if present) to observe and record the rehabilitation session. We did not document any identifiable patient data. The patients were not the focus of the field study; observations were primarily to get insight on the work practices of the rehabilitation specialists. Participants were not compensated for their participation.

% ========== ==========
\subsection{Data Analysis}
% ========== ==========
We performed a qualitative data analysis focusing on the \emph{information} used in rehabilitation and the needs for a telerehabilitation system. Two researchers analyzed the data, a PhD student (information scientist) with over 5 years of research and development experience in HCI; and an undergraduate student in mechanical engineering. One author also has experience as a caregiver. We compiled our field notes, transcribed all the observations and interviews, then performed open-coding of all data in three iterations, refining the codes as we progressed. The two coders (first and third authors) compared concepts informally as they coded, discussing differences in interpretation and reaching either a consensus or documenting their differences. 

After open coding, we used an inductive analysis to create categories based on behaviors exhibited by the participants. Our observation codes included: Personal Check In/Conversation (when the specialist prompts the patient), RS Describe Task(s)/Goals, Start Activity, End Activity, Rehabilitation Specialist Demonstrates Task with Body, Rehabilitation Specialist Asks Patient to Move a Certain Body Part, and Rehabilitation Specialists Taking Notes. After an initial round of coding, we noticed that Personal Check In/Conversation was the most-used code. Therefore, we focused on having a conversation with the specialists, which turned our analysis efforts towards the details on the conversation. Our final set of interview and observation codes included: Describes Caregivers Involvement, Information Documented During Session, Interdisciplinary Interaction, Medication Management Conversation, and Comorbidity Discussion. Lastly, the first author grouped codes and found relations among them to create the four high-level themes presented. There was one overarching theme that described all the data, experiential information about a stroke survivor plays an important role in rehabilitation. 

%
% ==================== ====================
\section{Experiential Information Is Essential within Stroke Rehabilitation}
% ==================== ====================
Our research study began under the assumption that the ideal telerehabilitation system optimally uses sensors to accurately measure movement and present it to a rehabilitation specialist. However, we quickly challenged this assumption, as our findings revealed that although movement data is an important component of stroke rehabilitation, it is not the only information need for specialists.
\emph{Experiential information} in stroke rehabilitation is information describing a stroke survivor's lived experience, collected through the \emph{patient-centered} approach and used to inform the rehabilitation care plan. Note, in our field study, multiple participants (\eg PT1, PHY2, OT6, OT7) used a synonym of patient-centered, \emph{client-centered}. 

Ultimately, experiential information is key to rehabilitation specialists' ability to build context around the patient's health status, and movement data. We will now detail how experiential information is built and it's components.
% ========== ==========
\subsection{Experiential Information Is Gathered Subjectively}
% ========== ==========
Rehabilitation specialists use a patient-centered approach to subjectively gather experiential information to build context. The approach includes strategies like conducting interviews during the rehabilitation session. It is important to understand that the different specialities gather experiential information and build context similarly, as they all use a patient-centered approach, but they prescribe and perform different interventions and evaluations.

We observed throughout our field study that the main job of specialists is not limited to objectively assessing stroke survivors, as OT2 commented during an interview: \textit{``You have to remember that our job is more than exercising their [patients'] arm.''} During rehabilitation sessions, we observed specialists used interviews to build context on their patients, OT3 explained: \textit{``Therapists first start with a subjective interview. They then ask them [stroke survivor] where their life situation is currently, who is looking after them, and what they can do for themselves.''} The experiential information is gathered from subjective interviews and directly influence the rehabilitation plan that specialists prescribe to their patients, and also helps specialists navigate the nuance and complexities of the stroke rehabilitation process. PHY1 said: \textit{``A number of underlying issues affect stroke rehabilitation. Physical, cognitive, social. A challenge is how to distill. There isn't always a set path, you have to play it by ear, you have to get the medical history and the physical[sic]. And the approach cannot always be solved algorithmically [sic].''}

\begin{figure}[htbp]
  \centering
  \includegraphics[width=1\linewidth]{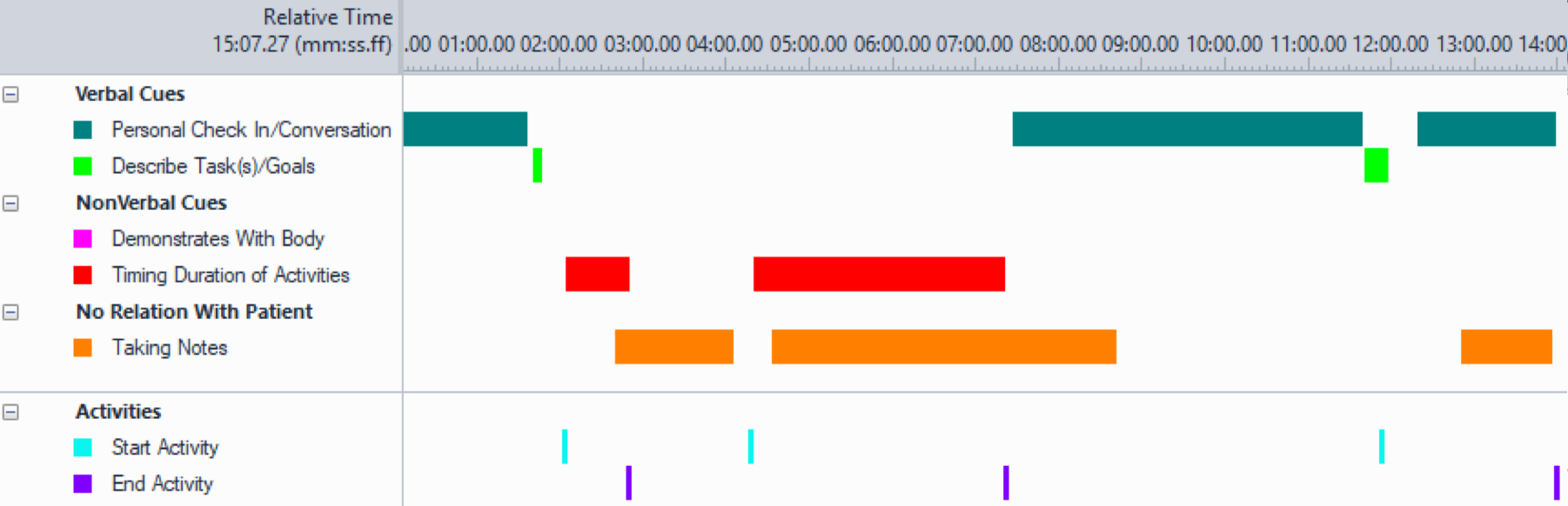}
  \caption{A 14-minute visualization of how time was spent during a rehabilitation session. The different activities are shown on the left, and bars represent their duration.}
  \Description{In this figure is a table timeline. On the y-axis is a list of events that occurred during a rehabilitation evaluation session On the x-axis is the time that has elapsed ranging from 0 minutes to 14 minutes.They're multiple colored bars that represent the categories the time in which that category occurred. The list of events on the y axis are: Personal Check In/Conversation (i.e., experiential information)(when the specialist prompts the patient), RS Describe Task(s)/Goals, Start Activity, End Activity, Rehabilitation Specialist Demonstrates Task with Body, Rehabilitation Specialist Asks Patient to Move a Certain Body Part, and Rehabilitation Specialists Taking Notes.}
  \label{fig:observation-time}
\end{figure}

Since the process is not built \textit{algorithmically}, specialists put the majority of their effort into gathering experiential information during rehabilitation sessions, when compared to gathering movement data. \autoref{fig:observation-time} (visualization made with Noldus ObserverXT \footnote{\url{https://www.noldus.com/observer-xt}}) shows a 14-minute snapshot of an example movement-based rehabilitation task taken from a 45-minute session. Here, the patient was required to complete a \emph{Nine Hole Peg Test} while the therapist observed. In a 14-minute span, the specialist acquired experiential information for $\sim$7 minutes (green) and took notes for $\sim$6.5 minutes (orange), whereas the actual activity lasted only $\sim$4 minutes (red). The remaining 30 minutes consisted of additional rehabilitation tasks, a medication management conversation, checking the patient's blood pressure, and creating new goals.

Rehabilitation specialists documented their subjectively-gathered experiential information by taking meticulous and systematic notes. One specialist shared with us a template used for session notes, based on the SOAP method. Here, specialists collect and document \emph{Subjective} data (\eg experiential information and context) and \emph{Objective} data during an appointment, to ultimately form an \textit{Assessment} and create a \textit{Plan}. Due to patient confidentially, our research team did not have access to patient medical notes, so we provide an example patient note that was created by Susan \etal~\cite{Susan2002} using the SOAP method (\autoref{app:SOAP}).

Objective information can also become experiential information through subjective contextualization. OT5 explained she puts her health assessment (\eg comorbidities) under the \textit{Subjective} section, including her professional medical opinion on how this affects physical abilities (\eg fatigue related to medication), as this impacts the care plan. In an interview with PHY3, he shared with us the importance of documenting subjective insight into his medical notes. PHY3 gave the example of a musculoskeletal evaluation, what the patient reports to him goes under the \textit{Subjective} section, and the results of the evaluation will go under the under \textit{Objective}.

% ========== ==========
\subsection{Stroke Survivor's Motivation, Stress, and Frustration Levels is Experiential Information}
\begin{figure}%[htbp]
  \centering
  \includegraphics[width=.67\linewidth]{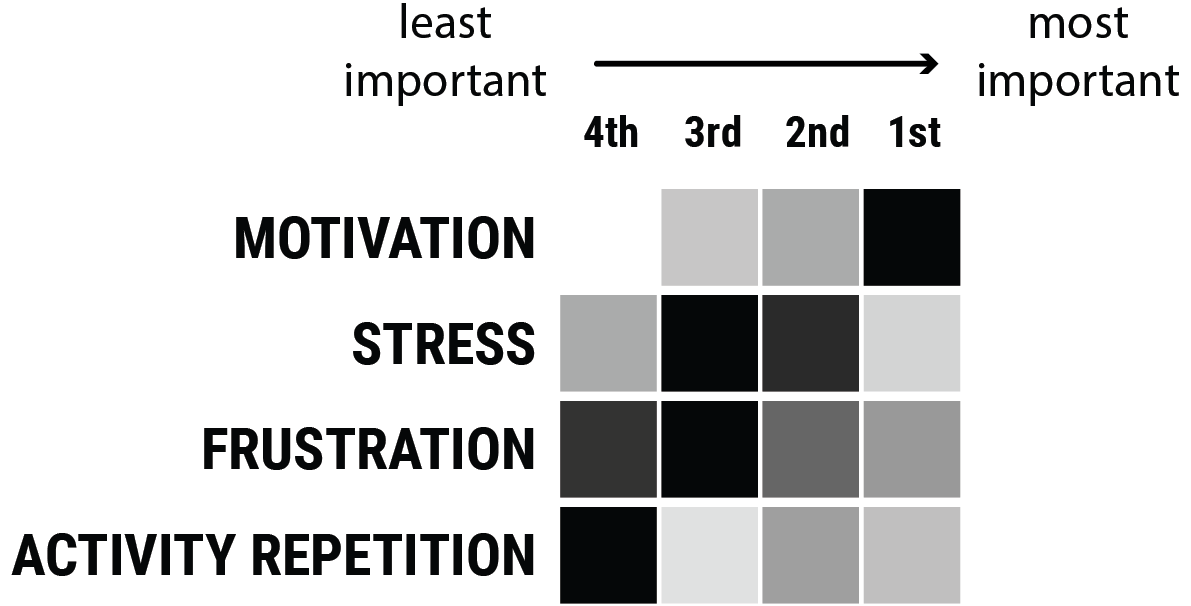}
  \caption{Heat map showing the ranking of importance of experiential information related to activity repetitions, stress, frustration and motivation. Visualization made using the Bertifier tool~\cite{6875988}}
  \Description{The heat map shows a trend where the majority of participants ranked as most important motivation, then stress, then frustration, and finally activity repetition.}
  \label{fig:ranking-chart}
\end{figure}

% ========== ==========
We found that factors like a stroke survivor's \textit{motivation, stress, and frustration levels} are a top priority for specialists to help gather experiential information. These levels are important because they all impact the ability for a stroke survivor to make progress on, and comply to, a care plan. We observed PHY4 explain this impact to her patient: \emph{``Things that make you feel like you're going down hill, or not recovering as fast, is depression, stress and fatigue. All of those things can make your symptoms feel a lot worse.''} In fact, when asked to rank the various types of information, 14 of the 16 participants ranked motivation, stress and fatigue as more important than activity repetition (\autoref{fig:ranking-chart}). All specialists expressed how important it is to gauge the patient's motivation level, as it informs the specialists' approach to prescribe a motivating and satisfying rehabilitation plan.

%\newpage

PT1 articulated how a rehabilitation plan is not bounded to objective assessments, but it has to be tailored to the motivations of the patient: \textit{``If they just do it [exercise] just once a day. Well, I am happy. What I try do is, because life is so on the go, make the rehab just part of their day. Like walking around the block.''} Simply, PT1 was sharing that the rehabilitation process is about creating a plan around the \emph{situated} conditions. PHY2 had similar sentiments as PT1: \textit{``Just getting a patient to exercise is good enough for me. I do not care about the reps.''} PHY2 shared examples of situating a care plan, such as assigning activities like walking around the neighborhood for patients that have a dog, or walking to a place she knows the patient will enjoy. The importance of the rankings can also vary because different factors dynamically change for each patient. PT2: \textit{``It [the order] depends on their home life, and things like that. Motivation is first because you don't get that, you don't get anything.''} PT2 shared similar sentiments, and said that her ranking would change for one of her patient's that has a stressful home. 

Stress levels are also included as experiential information because they can have a direct impact on the stroke survivor's physical health, such as blood pressure. OT7 expanded on the importance of also understanding the source of the stress such as \textit{family stress} and \textit{personal stress}, thus ensuring that prescribed rehabilitation plans do not exacerbate the patient's stress level, blood pressure and/or heart rate levels. PT4 shared, \emph{``I tend to look at the big picture, and I might give them a home program [rehabilitation plan] that doesn't stress their lives too much, just so they do it. I try to prioritize quality of life and compliance.''} She went further to expand on her role, and what she considers the big picture. PT4 said: \emph{``My role as a PT is to get them as independent as possible, and try to return them to fun things. For example, I tried to get him [the patient] back to golfing because it is a stress reliever for him and he enjoys it. I am particularly passionate to get them back to their normal role as much as possible''.} Including stress levels as experiential information allows specialists to gauge quality of life, and prescribe an appropriate care plan. 

Finally, frustration levels play an important role \emph{during} prescribed rehabilitation exercises and tasks. While observing OT6 conduct a rehabilitative exergame with her patient, she informed us about her process of managing frustration during activities: \textit{``We have to have an idea of the game, and then we pair it with their [patients] level of function. We give them a challenge, but not so much that they can't be successful, otherwise it is just frustrating.''} What we see here is the specialist paying more attention to rising frustration levels than focusing on the movements themselves. Her goal is not to achieve a certain number of movements, but instead she is looking for the right balance of effort against frustration. Thus, frustration levels are experiential information that better inform specialists, so they prescribe attainable activities. Simply, having context on what motivates, stresses, and frustrates the stroke survivor is crucial for both devising an exercise plan, as well as assessing the survivor's success at following the plan. It is more important for the specialists to learn if there is a moderate level of motivation, low number of stressors, and low frustration than to gather any further information on what movement the patient actually performed. 

% ========== ==========
\subsection{Stroke Survivor's Acclimation \& Goals are Experiential Information }
% ========== ==========
The survivor's \textit{acclimation} after a stroke and their rehabilitation \textit{goals} are considered experiential information. A survivor's acclimation is simply how they are coping in their home after surviving a stroke; for example what kind of accessibility or health related issues they are having. Goals are a set of health milestone patients want to achieve through rehabilitation, such as returning back to work or driving for themselves. These two aspects are tightly connected because a survivor's acclimation can inform the progress made towards accomplishing a goal. 

To evaluate the patient's acclimation at home, specialists ask questions to determine if their patients are successfully preforming Activities of Daily Living (ADLs). ADLs are basic daily self care tasks (\eg managing medication, meal planning, or shopping) Typically OTs are the specialists that assess ADLs~\cite{KATZ1983} as a data point, but PT4 mentioned that she also references ADLs with her patients. PT4 said, \textit{``I do blend some with OT on the ADLs. Like I work on toileting with patients. The OT might work on using assistive device, but I do more of the gross motor stuff. Like can you transfer yourself from the wheelchair to the toilet.''} PT4 essentially uses the patient's experience on ADLs as a reference to inform the type of intervention she prescribes towards a goal.

It is important to evaluate acclimation, as this helps specialists create or modify the care plan while staying aligned to the patient's goals (\eg dressing independently). However, it is more complex than simply asking if, when, or how exercises were completed, experiential information is required.
One strategy of evaluating and validating the acclimation is using sensor data to probe further about the patients' experience at home. PHY2 for example asks some of her patients to wear a \emph{Fitbit}\footnote{\url{https://www.fitbit.com/}}, so then she can review the data during the rehabilitation sessions. PHY4 recalled a time when a patient step count steadily declined, causing her to inquire if the change in activity was due to a comorbidity or decreased motivation, a crucial difference when evolving the care plan. Validating home exercises is important for specialists to determine if rehabilitation plan goals are being met. Validation can also be performed by patients themselves through self-reporting, as OT4 explained: \emph{``Validating home exercises is a combination of objective and subjective measures [..] I actually tell my patients to complete a weekly report for me''.}

Anticipating acclimation is necessary to adapt goals before it is too late, so specialists use contextual information about a patient's schedule.
For example, during an observation, the stroke survivor and caregiver informed PT2 that the survivor was having their family over later that evening for their birthday.
In response, PT2 reduced the exercise for her patient because she, \textit{``wanted him to survive tonight.''} 
Taking into account what the patient's acclimation will be later on (tired), she altered the rehabilitation plan for that particular session, recognizing that the goal for the evening was to enjoy time with their family.

% ========== ==========
\subsection{Stroke Survivor's Health is Experiential Information}
% ========== ==========
\emph{Physical} and \emph{mental} health are key experiential information when prescribing a rehabilitation plan, as both impact the ability to complete the plan. Additionally, mental health can impact compliance.
% ===== =====
\subsubsection{Physical Health}
% ===== =====
%%
% Comorbidities are physical health
%%
Multiple comorbidities are common in stroke survivors. Some examples we heard during discussions include: hypertension, diabetes, malnutrition, sleep apnea, depression, insomnia, pseudo dementia, atrial fibrillation and vertigo. Comorbidities require close monitoring, management, and coordinated care amongst the stroke survivors coordinate care amongst the care network because they dictate how successfully and safely exercises are performed. Typically, these are managed through situational changes, such as changing diet or minimizing stressful activities, and medication.
 
\begin{figure}[htbp]
  \centering
  \includegraphics[width=1\linewidth]{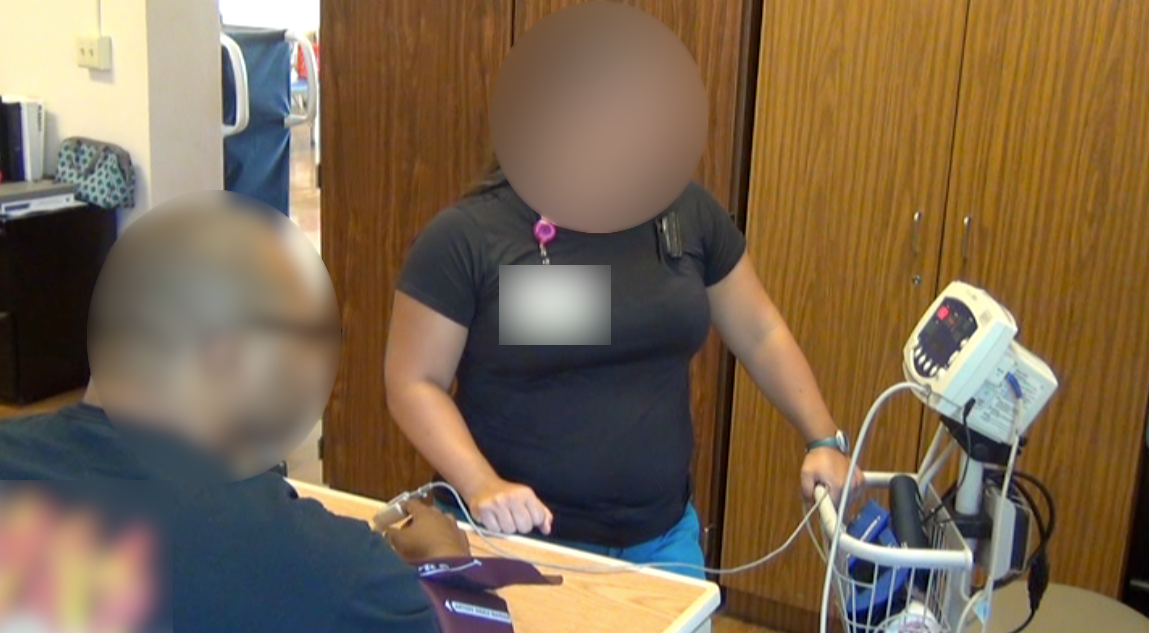}
  \caption{Specialist taking the patient's blood pressure after an activity.}
  \Description{In this figure a patient is getting their blood pressure read by the rehabilitation specialist. The specialist is standing, and the patient is sitting. The patient appears to an African-American male, and the specialist appears to be a white woman.}
  \label{fig:rp-bp}
\end{figure}

Some of the participants, such as PHY1, PHY4 and OT1, took a patient's vital signs at the end of an activity to check for fatigue (\autoref{fig:rp-bp}), determining the situational changes for the exercise program. For example, in an observation with OT1, she noticed signs of discomfort and fatigue, so she took vitals and recorded them in her medical notes, ending the exercise early. In the followup interview, she told us that this patient's hypertension limits their ability for certain physical and/or cognitive interventions.
\newline
\newline

%\newpage

% ===== =====
\subsubsection{Mental Health}
% ===== =====
Specialists have a particular interest in understanding cognitive abilities and depression, when inquiring about mental health.
%%
% Cognitive ability
%%
To assess level of cognitive ability, we observed how PHY4 and OT1 refer to the MOCA. PHY1 and PT3 later explained during interviews that such assessment is important in determining what the patient can do, thus impacting the care plan, but also in understanding the patient's own ability to comprehend the exercises. In the cases where cognitive impairments hinder compliance, the specialists will often coordinate with a caregiver, so they can assist the stroke survivor with completing prescribed interventions while at home. 

\newpage

Depression in particular can impact the patient's motivation to comply with the prescribed rehabilitation plan while at home. In our observations, we observed physiatrists (PHY1 and PHY4) speak extensively with stroke survivors and their caregivers about the survivors' ongoing battle with depression, and then prescribed interventions. In the observation with PHY4, she modified the patient's rehabilitation plan by first reaching out directly to the patient's neuropsychologist to discuss and coordinate a response to the patient's bout with depression. 

% ========== ==========
\subsection{Caregiver's Assessment is Experiential Information}
% ========== ==========
Many specialists considered the rehabilitation process as a team effort, which includes the caregiver.
Caregivers regularly assess the patient, which becomes important experiential information. They are deeply involved with multiple aspects of a stroke survivor's life, they can be a family member, friend, or hired professional. Their role is so important in the survivor's recovery process that PT3 went as far as saying, \textit{``I think caregiver support is a major predictor into how much someone can recover. Fortunately, they can offer support, encourage [them], [overcome] cognitive deficits, [monitor] their schedule, [help with] exercise, and [cook] meals. The biggest area is compliance.''}

Caregivers play an important role in the care plan, both facilitating and validating the plan. For example, in a session with PHY4 the patient self-reported that they felt they were speaking much better. However, the caregiver felt differently, and provided a more in-depth assessment of the progress: the patient's speech indeed has improved, but slowly began to decline in the weeks leading up to the appointment. PT4 expanded on the value of this assessment: \emph{``It is pretty nice to get the caregiver's perspective, because they're a little bit more honest than the stroke survivor. I think they are really important, because they encourage the patient to get better, and they are the key for their patient live their life again. I like them to be in the therapy session.''}

Having the caregiver present during a rehabilitation session is highly valued because they provide insight on the patient's experience at home and offer assistance. We observed caregivers providing assistance to the specialists (\eg supporting the stroke while walking), including insight when the specialists are completing an assessment. OT4 elaborated on how the experiential information from the caregiver complements the patient information: \emph{``Sometimes people [rehabilitation specialists] will give out an assessment to the patients and caregivers. That way, we will get the patient's insight on how they performed different ADLs, and we will get the family members' insight on how they felt the patient performed the ADLs.''} In this situation, the family member is reporting not on the exercise \emph{execution}, but on how the patient is \emph{faring} in their daily lives---what they are able to do and how they are doing it. This constitutes key experiential information.

\newpage

% ==================== ====================
\section{Discussion}
% ==================== ====================
We found that experiential information is essential in co-located stroke rehabilitation for rehabilitation specialists, and that they use a large variety of information other than exercise movement data.
However, the focus of home-based therapy systems is on movement data, as they aim at motivating (\eg Us'em~\cite{Beursgens:2011:UMS:1979742.1979761}) and monitoring patients (\eg ArmSleeve~\cite{Ploderer2016}).
This means that they (1) present movement data to specialists upfront and center, (2) have not enabled capturing experiential information of the movement data, and (3) do not provide means for annotating movement data for context. This has led to a paradigm that puts emphasis on movement data (\autoref{fig:paradigm-reconceptualization} - left), and, as a consequence this legacy has led telerehabilitation systems to focus on the computational work for recognizing movement and quantifying it for remote specialists to visualize (\eg TeleREHA~\cite{Perry2011}, mRes~\cite{Weiss:2014:LCT:2686893.2686989}, Rehabilitative TeleHealthCare ~\cite{Postolache2011}, and systems covered by Santayayon \etal~\cite{Santayayon2012}).
Our study shows that this paradigm is incongruent with how specialists actually work in face-to-face stroke rehabilitation. This does not invalidate the motivation in the current paradigm to track movement and count repetitions accurately for telerehabilitation. Instead, we posit that movement data needs to exist within a more sophisticated understanding of patients' experiential information, when designing telerehabilitation systems.

% ========== ==========
\subsection{Paradigm Reconceptualization}
\begin{figure*}[htbp]
  \centering
  \includegraphics[width=1\textwidth]{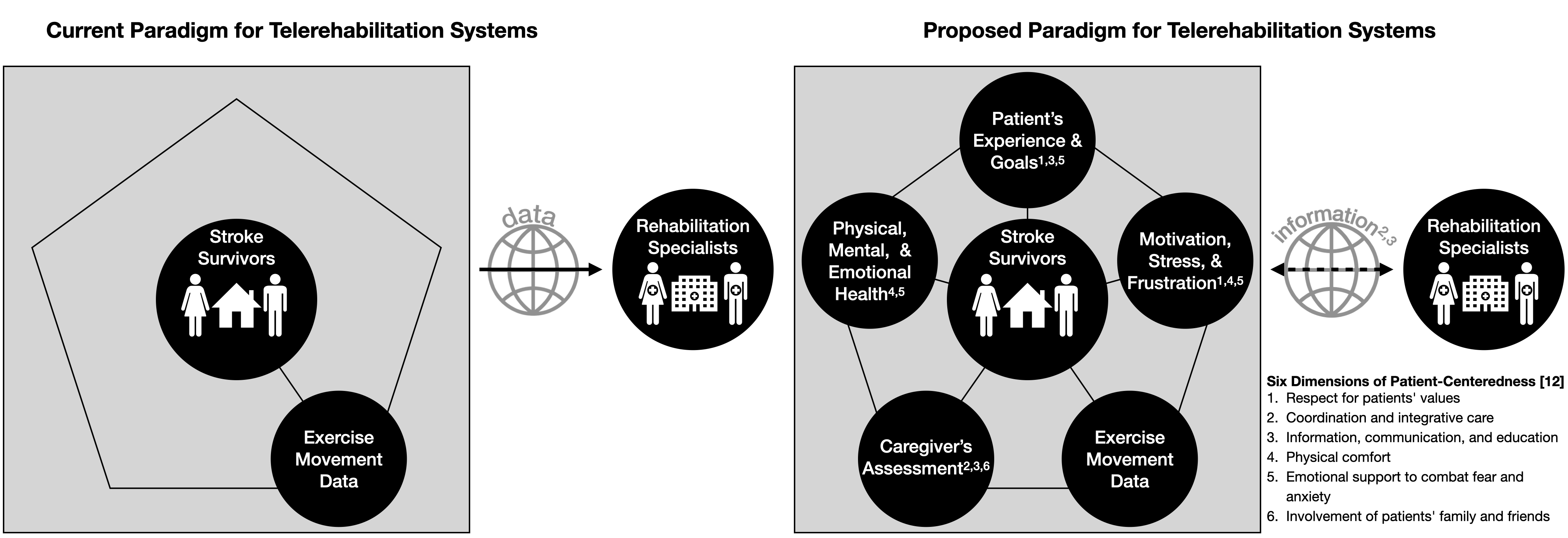}
    \caption{Current Paradigm for Telerehabilitation Systems}
    \Description{This figure shows two paradigms designed by the authors, one on the left (labeled “Current paradigm”) and one on the right (labeled ”Proposed Paradigm). All the way to the right, there is a list of six dimensions of patient-centeredness [[reference number 12], including 1. Respect for patients' values, 2. Coordination and integrative care, 3. Information, communication, and education, 4. Physical comfort, 5. Emotional support to combat fear and anxiety and 6. Involvement of patients' family and friends. The left “Current telerehabilitation paradigm” has three main components. The first component, on the left side, is a pentagon. This pentagon represents a telerehabilitation system at the home of the stroke survivor. At the center of the Pentagon, is a circle that is labeled stroke survivor, and there is a solid line connecting the stroke survivor’s circle to a circle at the bottom right point of a pentagon. This additional circle at the bottom right is titled "exercise movement data". The other four points of the Pentagon are empty. The second component, in the middle, is an icon of the earth with a solid one way arrow cutting through the middle of the earth. The arrow is pointing from the first component to the third component. The second component represents the one way flow of "data" over the Internet. The third component, on the right side, is a circle that says "rehabilitation specialists". This third and final component represents rehabilitation specialists receiving data at their respective medical center or working location. In summary, the current paradigm of telerehabilitation represent data that is collected at home is exercise movement data, and that data is then sent one way over the internet to the rehabilitation specialists. It is important to note, that the data does not flow backwards from the third component to the first component. This is important because the authors’ proposed paradigm has a two way communication. The authors’ proposed telerehabilitation paradigm reconceptualization. There are three main components to this figure. The first component, on the left, is a pentagon. This pentagon represents a telerehabilitation system at the home of the stroke survivor. At the center of the Pentagon is a circle that  is labeled stroke survivor, and there is a solid line connecting the stroke survivor’s circle to five circles that are revolving around the stroke survivor center circle. The five circles are at the five points of the pentagon. The five circles represent the five aspects of experiential information, and they are titled, "Exercise movement data", "Motivation, Stress, & Frustration", "Patient’s Acclimation & Goals", "Physical, Mental & Emotional Health", "care givers' assessment". This propose paradigm is different from the current paradigm because there are 5 data points that are revolving around the stroke survivor, compared to the current paradigm that only has one (i.e. Movement data). The second component, in the middle of the figure, is an icon of the earth with a dotted two way arrow cutting through the middle of the earth. The arrow is pointing between from first component to the third  component. The second component represents the two way flow of "information" over the Internet. The third component, on the right side of the figure, is a circle that says "rehabilitation specialists". This third and final component represents rehabilitation specialists receiving information at their respective medical center or working location, but also sending back information . In summary, this figure represent the author’s proposed paradigm of telerehabilitation. Simply, experiential information is collected at home, that information is then sent over the internet to the rehabilitation specialists, and the specialists have the opportunity to send back relevant information to the stroke survivor (i.e. Two way dialogue) . It is important to note, that the information does flow backwards from the third component to the first component. The last important aspect of this figure is on the bottom the right of this figure, there is a list of six dimensions of patient-centeredness [reference number 12]. Again, The six dimensions are 1. Respect for patients' values, 2. Coordination and integrative care, 3. Information, communication, and education, 4. Physical comfort, 5. Emotional support to combat fear and anxiety and 6. Involvement of patients' family and friends. The proposed paradigm further illustrates how the authors’ paradigm integrates the six dimensions of patient centeredness into the five aspects of experiential information. The connections are: "Motivation, Stress, & Frustration" integrate 1. Respect for patients' values, 4. Physical comfort, 5. Emotional support to combat fear and anxiety , "Patient’s Acclimation & Goals" integrate 1. Respect for patients' values,3. Information, communication, and education,5. Emotional support to combat fear and anxiety , "Physical, Mental & Emotional Health" integrate 4. Physical comfort, 5. Emotional support to combat fear and anxiety, "care givers’ assessment" integrates 2. Coordination and integrative care, 3. Information, communication, and education, and 6. Involvement of patients' family and friends. In sum, this figure illustrates that the author’s paradigm incorporates experiential information that integrates patient centeredness, and encourages information to be shared back and forth between the stroke survivor and their rehabilitation specialists.}
    \label{fig:paradigm-reconceptualization}
\end{figure*}

% ========== ==========
Previous work hints towards the need to capture information beyond movement.
For example, How \etal~\cite{How2017c} stated that successful rehabilitation systems should adapt to surrounding life context, and Ploderer \etal~\cite{Ploderer2016} recognized that lack of contextual information hinders the interpretation of movement data. The two studies however do not go into detail defining or modeling what is ``context'' information beyond movement exercise data. Through our work, we studied such information needs, and this has enabled us to move forward the paradigm through which systems are designed.
We propose a paradigm reconceptualization for \emph{situated} telerehabilitation systems (\autoref{fig:paradigm-reconceptualization} - right ), that capture the experiential information used in rehabilitation: a stroke survivor's (1) motivation, stress, and frustration levels, (2) acclimation and goals, (3) their health status, and (4) the caregiver's assessment; where exercise movement data complements the experiential information through a semantic layer that gives meaning to otherwise incomplete data. In this way, the sharing of a stroke survivor's lived experience within their care network is now taken into consideration.
This brings telerehabilitation system design closer to how co-located rehabilitation takes place, by integrating the patient-centered approach~\cite{gerteis1993} and giving meaning to movement data by incorporating experiential information on the stroke survivor's lived experience.

\newpage
% ===== =====
\subsection{Implications for Design}
% ===== =====
Below we provide design implications to capture context by leveraging our findings that identified types of experiential information.

\subsubsection{Capturing Experiential Information to be Congruent with Rehabilitation}
For telerehabilitation systems to be congruent with the existing rehabilitation practices, they need to provide ways to capture and share various forms of experiential information.
\textsc{Motivation, Stress, and Frustration}. The insight we received from specialists during the interviews (\eg \autoref{fig:ranking-chart}) reveal the importance of motivation, stress, and frustration, and that keeping track of the factors on a regular basis can help trace the cause for changes in exercise movement and daily activity. Capturing a patient's motivation, stress and frustration levels is essential for specialists to build life context for their patients. Capturing the three could look like a system autonomously prompting questions to the patient or caregiver to answer when deviations in any of the three factors are detected. Capturing this information in the moment would allow contextualization of the patient's exercise/movement data and overall experience. The prompts will ideally elicit responses that inform the specialists the reason behind the deviation. Additionally, specialists could predetermine alternative exercises and activities they know are engaging for a particular patient, and the system could automatically propose the appropriates activity to respond to the contextual reasons underlying change in movement. 

\textsc{Acclimation \& Goals}. Telerehabilitation systems would benefit from including features that capture acclimation and goals to help specialists evaluate patients' progress. How \etal~\cite{How2017c} briefly discussed how there may be better ways in which telerehabilitation systems can seek to complement rehabilitation goals of traumatic brain injury (\eg stroke), but did not offer concrete design suggestions. To benefit acclimation and goal tracking, systems can implement a feature that allows patients to create a profile or virtual diary that acts as a virtual acclimation report. This report would capture how a patient is getting acclimated to a new environment and everyday tasks, that is accessible to their care networks.
For example, data from a smart watch or \emph{Fitbit} could be used to populate a home acclimation report instead of being used simply as a measure of exercise. The \emph{Fitbit} movement data can be tagged either as performing prescribed exercises or simply ADLs in the report. In this way, an event such as a decrease in tagged ADL movement data would signal that the survivor is probably not acclimating well. Additionally, the specialist would use this report to discuss the patient's goals immediately or during a session.

\textsc{Physical Health.} There have been examples of systems that collect exercise movement data for physical health, TeleREHA~\cite{Perry2011} monitors arm reach and mRes~\cite{Weiss:2014:LCT:2686893.2686989} monitors wrist function. However, our study showed that physical health is not limited to exercise movement data. A telerehabilitation system that acquires both physiological and exercise movement data brings additional benefits when working with survivors with comorbidities (\eg diabetes and hypertension).
By periodically checking heart rate (\eg via smart watch) and or blood pressure (\eg via a built-in blood pressure monitor), a system can alert the appropriate specialist when these levels reach concerning thresholds in combination with movement data during rehabilitation exercises. This would enable specialists to intervene remotely and modify the rehabilitation plan based on accurate information, for example suggesting a rest. In a synchronous telerehabilitation system, the specialist might suggest a rest. In an asynchronous system, the data could be used to start discussion at a later time.
Rehabilitative TeleHealthCare~\cite{Postolache2011} was the one telerehabilitation system we found that aligned with our finding to combine physiological data and movement data. TeleHealthCare uses sensors to capture physiological data (heart rate, oxygen saturation (SpO2) and respiration rate) and movement data for health monitoring. However, this system does not take into account what are the patient comorbidites (\eg Diabetes and Atrial Fibrillation), and thus limits how an expert can use these data combined.

\textsc{Mental Health.} As our study showed, mental health is another aspect that impacts the creation of a rehabilitation plan, and its compliance given a patient's cognitive function. A system that leverages existing standardized assessments (\eg MOCA) to collect data will better match current rehabilitation we observed. To keep track of the response evolution of such assessments, a telerehabilitation system can be used to require patients to complete cognitive assessments when prescribed by the specialist throughout time. It is important that systems collect, keep track of, and transmit mental health status of the patient to specialists, as this information is an experiential data point used to contextualize deviations in exercise movement data, along with motivation stress, and frustration levels. 
Moreover, it is particularly important to consider the collaboration among the care network when it comes to health data and sticking to existing workflows.
As Ng \etal~\cite{Ng2019} recently stressed, health care providers are concerned about how to adjust their practices, to better guide the use of sensor data within mental telehealth.
By keeping all the care network updated with mental health information, a telehealth system can enable the care network to collaboratively interpret the collected exercise movement and mental health data, avoiding misinterpretation, but also allowing for shared decision making when it comes to adjusting how sensor data is used.

\textsc{Caregiver Assessment}. As we learned, caregivers provide crucial support for a successful execution of a rehabilitation plan at home, and thus are the ones that hold key information about the day-to-day issues around the plan. However, no system to our knowledge considers incorporating information from a caregiver. Telerehabilitation systems should incorporate functionalities that capture caregivers' insights around the execution of exercises, so that specialists can compare and contrast this view with both the patient's view on their own execution, and the movement data. More importantly, they can act as surrogates for filling reports or reporting issues, when the stroke survivor is not able to do so, such as when the survivor's health is negatively impacted by a comorbidity. In this way, the rehabilitation specialist can better evaluate the health progress of the stroke survivor.
% ===== =====
\subsubsection{Annotating Experiential Information to Create Shared Meaning}
% ===== =====
One of our main insights is that specialists create meaning subjectively. We observed strategies around creating meaning on top of movement data, such as reviewing \emph{Fitbit} data together with a specialist. Ploderer \etal~\cite{Ploderer2016} concluded in their study that a major limitation of \emph{ArmSleeve} is the lack of contextual information presented across the different dashboard pages; and Mentis \etal~\cite{Mentis2017} showed that clinicians require patient contextual information to make sense of \emph{Fitbit} data---for instance if a high walking day was due to a vacation and a low walking day was due to a poor medication reaction. Telerehabilitation systems should have a dedicated semantic information layer for all the information stored. It is likely that different specialists have a unique perspective on data and want to annotate different data (\eg the data related to their own profession). Having these annotations can lead to less confusion about what other specialists are doing, how they deal with issues, and most importantly, it will facilitate coordination.

% ===== =====
\subsubsection{Supporting Coordination to Consolidate the Care Plan}
% ===== =====
Related to our last point, we believe that telerehabilitation systems have the potential to be the central hub of information around rehabilitation. We observed the level of care coordination involved between those in the stroke survivor's care network. Therefore, it is important that telerehabilitation systems provide functionalities that best facilitate the communication and coordination of the different specialists by allowing them to access each other's data, add their own interpretation, and thus coordinate their actions. We assert this can play a major role in the care plan, as this is the one artifact where all the actors have an influence, making it dynamic.
Accommodating for dynamic, always up-to-date care plans with transparent decision making imprinted on the plan itself (through annotations), is a way to keep care consolidated.

% ========== ==========
\subsection{Limitations \& Future Work}
% ========== ==========
Our work is limited to function and mobility rehabilitation specialists within the Physical Medicine \& Rehabilitation specialty. This work does not include the other specialists that are involved in the rehabilitation processes of stroke survivors or other illnesses that require rehabilitation. Now that we have working relationships with our partners, we will conduct observations and interviews with more rehabilitation specialists. We will also begin eliciting feedback from additional stakeholders (stroke survivors and caregivers) on our paradigm refocus. Ultimately, with an eye to begin developing a prototype of a situated telerehabilitation system.

% ==================== ====================
\section{Conclusion}
% ==================== ====================
Our research reveals that experiential information is an essential need in stroke rehabilitation. This paper provides a more holistic view of the practices of rehabilitation specialists within the Physical Medicine \& Rehabilitation medical specialty. Additionally, this paper proposes a paradigm reconceptualization in stroke telerehabilitation system development to address the complex and dynamic nature surrounding stroke rehabilitation. We posit that our proposed refocus can lead to the development of telerehabilitation systems that will be on par with the type of interactions and evaluations that exist in a face-to-face stroke rehabilitation session. We do not suggest that sensors have no role in telerehabilitation, but we say to our research community, that the exercises are a means to the end, and it is important to understand what the end is.

\begin{acks}
We would like to thank the patients and medical personnel that participated in our study. We would like to thank Dr. Wittenberg for their insight and assistance. This work has been supported in part by NIH/NIGMS R25 GM055036 IMSD Meyerhoff Graduate Fellows Program, NSF CAREER award IIS-1552837 and SaTC award CNS-1714514, NIH/NIGMS MARC U*STAR T34 08663 National Research Service Award, and by Dean Keith J Bowman and the UMBC Constellation Professorship.
\end{acks}
\newpage

\bibliographystyle{ACM-Reference-Format}
\bibliography{telerehabilitation.bbl}

\appendix
\section{Sample SOAP Note}
\label{app:SOAP}
\begin{figure}[htbp]
  \centering
  \includegraphics[width=.75\linewidth]{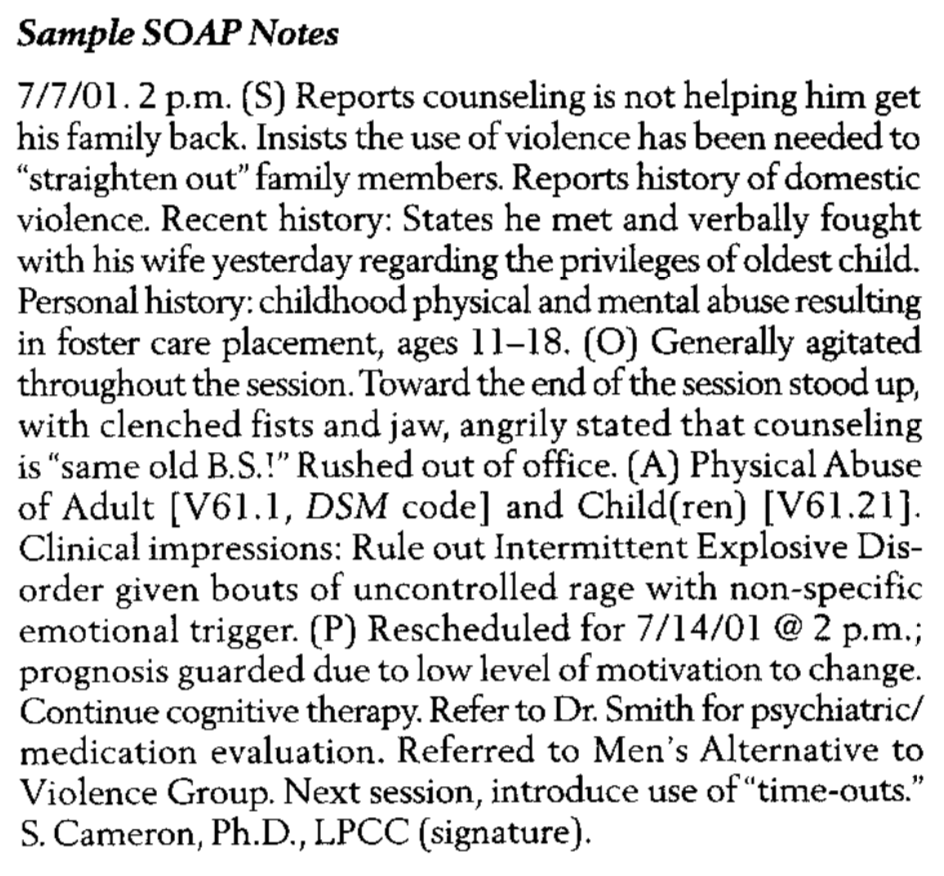}
  \caption{Example Medical Note using the SOAP method \cite{Susan2002}}
  \Description{This is screenshot of a sample soap note. The soap note says July 7, 200 and1 2 PM Subjective- reports counseling is not helping him get his family back. And sister use of violence has been needed to straighten out family members. Reports history of domestic violence. Recent history: states he met and verbally fault with his wife yesterday regarding the privileges of the oldest child. Personal history: childhood physical and mental abuse resulting in foster care placement, ages 11–18. Objective- Generally agitated throughout the session. Toward the end of the session stood up, with clenched fist and jaw, angrily stated that counseling is "some old BS!" Rushed out of the office. Assessment- Physical abuse of adult [ V6 1.1, DSM code] clinical impressions: rule out intermittent explosive disorder given bouts of controlled rage with non-specific emotional trigger. Plan – rescheduled for July 14, 2001 at 2 PM; prognosis is guarded due to low level of motivation to change continue cognitive therapy. Refer to Dr. Smith for psychiatric/medication evaluation. Referred to men’s alternative to violence group. Next session, introduce use of "time outs." Signed. S. Cameron, PH.D.LPCC}
  \label{fig:soap-example}
\end{figure}

In \autoref{fig:soap-example}, the medical specialist documented the patients comments (\emph{Subjective}) and physical actions (\emph{Objective}) to inform the assessment and prescribed plan.

\end{document}